# Probing the Time Domain with High Spatial Resolution

*Science White Paper for the Astro2020 Decadal Survey*

**Thematic Areas:**
- ✔ Resolved Stellar Populations and their Environments
- ✔ Cosmology and Fundamental Physics
- ✔ Multi-Messenger Astronomy and Astrophysics

**Authors:**

J. P. Blakeslee[1] (Gemini Obs.), S. A. Rodney (U. South Carolina), J. M. Lotz (Gemini), G. Sivo (Gemini), S. Sivanandam (U. Toronto), M. Andersen (Gemini), R. Carrasco (Gemini), L. Ferrarese (NRC Herzberg), R. J. Foley (UC Santa Cruz), S. Goodsell (Gemini), P. Hirst (Gemini), J. B. Jensen (Utah Valley U.), P. L. Kelly (U. Minnesota), A. A. Kaurov (IAS), M. Lemoine-Busserolle (Gemini), B. W. Miller (Gemini), J. O'Meara (Keck), H. Roe (Gemini), M. E. Schwamb (Gemini), J. Scharwächter (Gemini)

**Abstract:**

Two groundbreaking new facilities will commence operations early in the 2020s and thereafter define much of the broad landscape of US optical-infrared astronomy in the remaining decade. The Large Synoptic Survey Telescope (LSST), perched atop Cerro Pachón in the Chilean Andes, will revolutionize the young field of Time Domain Astronomy through its wide-field, multi-band optical imaging survey. At the same time, the James Webb Space Telescope (JWST), orbiting at the Sun-Earth L2 Lagrange point, will provide stunningly high-resolution views of selected targets from the red end of the optical spectrum to the mid-infrared. However, the spatial resolution of the LSST observations will be limited by atmospheric seeing, while JWST will be limited in its time-domain capabilities. This paper highlights the scientific opportunities lying between these two landmark missions, i.e., science enabled by systems capable of astronomical observations with both high cadence in the time domain and high resolution in the spatial domain. The opportunities range from constraining the late phases of stellar evolution in nearby resolved populations to constraining dark matter distributions and cosmology using lensed transient sources. We describe a system that can deliver the required capabilities.

---

[1] Contact author: jblakeslee@gemini.edu

**Scientific Context**

Early in the 2020s, two revolutionary new astronomical facilities with open access to the US Community should see their first light. First, the Large Synoptic Survey Telescope (LSST; Abell et al. 2009; Ivezić et al. 2018), funded mainly by the NSF and DOE, will begin a ten-year mission to produce a deep, wide, multi-band photometric map of approximately half the sky. Although the observing strategy is still being finalized, the proposed "universal cadence" is expected to yield $10^7$ transient alerts each clear night, ushering in a golden age for Time Domain Astronomy. The other open-access facility expected to revolutionize the optical-IR astronomical landscape is the James Webb Space Telescope (JWST), funded primarily by NASA. JWST's four instruments will provide imaging and spectroscopy from 0.6 to 27 μm. The imaging performance is optimized for 2 μm, where JWST achieves a diffraction-limited PSF with a FWHM of ~ 0.07" and the NIRCam short wavelength channel provides Nyquist sampling.

The two facilities will be complementary. LSST will be an optical imaging survey more sensitive and prolific than any previous one, but still limited by atmospheric seeing; the median delivered seeing is expected to be ~ 0.8″ in the red bands. Likewise, JWST will far exceed the performance of any previous facility for NIR imaging, and it will provide a variety of $R \leq 2700$ spectroscopic options, including multi-object, slitless, and integral field spectroscopy. However, although NIRCam's time-series mode will enable rapid photometric and spectroscopic monitoring of relatively bright sources, JWST's time-domain follow-up capability is quite limited because of slew times and pointing constraints. For targets within 45° of the ecliptic plane (70% of the sky), there will be two visibility windows per year centered six months apart. In the ecliptic plane, each window lasts 50 days, meaning that the target is unobservable for two stretches totalling almost 9 months. The observability increases with ecliptic latitude, but even 40° from the plane, the windows are each 75 days, so that the target would be unobservable for over 7 months during the year. Moreover, the current plan is to limit the number of rapid target of opportunity observations for JWST to just six per year, with a *minimum* turnaround time of 48 hours.

Thus, neither of these facilities can provide high-cadence long-duration monitoring of variable phenomena at very high spatial resolution. A multi-conjugate adaptive optics (**MCAO**; Rigaut & Neichel 2018) system, providing a corrected field over a wide (> 1′) area, and fully integrated into a queue-operated 8m-class telescope feeding both a well-sampled imager and a multi-IFU spectrograph, would provide this capability. No such facility exists. Although there is one MCAO system in use (GeMS at Gemini South; Rigaut et al. 2014), it is a high-maintenance system scheduled only for brief periods that limit time-domain work. A robust queue-operated MCAO facility capable of delivering JWST-like resolution over a NIRCam-like field on any suitable night would have great synergistic potential with JWST and open a variety of exciting opportunities in the time domain, as we discuss in this white paper. It would also provide superb synergy with WFIRST later in the 2020s and serve as a pathfinder for the more complex NFIRAOS (Herriot et al. 2014) MCAO system that will feed the TMT science instruments

**Stellar Evolution from Pulsating Stars in Metal-rich Environments**

Stellar evolution during the thermally pulsing asymptotic giant branch (**TP-AGB**) remains one of the major problems in modeling the panchromatic spectral energy distributions of galaxies. During the TP-AGB stage, stars undergo rapid mass loss, enriching the interstellar medium with the churned-up products of their nucleosynthesis and drastically altering their subsequent evolution (e.g., Herwig 2005; Marigo et al. 2017). Stellar population models predict that TP-AGB stars contribute a substantial fraction of the bolometric light at intermediate ages (e.g., Conroy et al. 2015; Figure 1); thus, an accurate understanding of the TP-AGB phase is essential for deriving mean ages and stellar mass-to-light ratios of galaxies, especially at high-redshift (e.g., Kriek et al. 2010; Rosenfield et al. 2016). Long-period variables (**LPV**s) are luminous stars with periods from ~ 80 to ~ 1000 days, believed to be evolving through the TP-AGB stage. They have been studied in many Local Group galaxies (Dalcanton et al. 2012; Boyer et al. 2016), and out to the Whirlpool Galaxy at 8.6 Mpc (Conroy et al. 2018). However, *there is no direct calibration of LPV luminosities or lifetimes in the metal-rich populations near the centers of massive early-type galaxies*. The main problem has been one of spatial resolution: it has not been possible to resolve individual LPVs within the crowded metal-rich regions of giant ellipticals.

Taking a novel approach, Conroy et al. (2015), demonstrated that the expected time-variable "pixel shimmer" due to luminous LPVs in Hubble Advanced Camera for Surveys I-band images of the M87, the massive cD galaxy in Virgo, can be detected in a region of the galaxy with [Z/H] > 0 dex. Interestingly, this study found that the inferred lifetimes of the metal-rich LPVs in M87 are 30% shorter than predicted by the models, implying larger than expected mass losses during the thermal pulsations. However, although the predicted statistical "shimmer" was clearly detected, the interpretation was model-dependent because individual LPVs were not detected. Moreover, the Hubble data spanned only 72 days, much less than the typical LPV period of ~ 300 days; a larger baseline is needed to constrain the higher luminosity slow pulsators.

With a facility capable of delivering NIR imaging with FWHM ~ 0.07″ over a > 1′ field on any clear night, it becomes possible to detect directly thousands of TP-AGB stars within single pointings on high-metallicity regions of the nearest giant ellipticals. This has been demonstrated with GeMS (Ferrarese 2018) but as noted above, this system is not well suited for temporal monitoring. Since LPVs are especially luminous in the *K* band, and time-series observations could be conducted regularly for a year or more, direct detections in the crowded regions of ellipticals as far as Virgo should also be possible. The measured numbers, luminosities, and periods of the LPVs would strongly constrain the TP-AGB lifetimes and contributions to total galaxy luminosities, essential for modeling the mass-to-light ratios and evolution of metal-rich stellar populations, especially in the NIR where models show the greatest disparities.

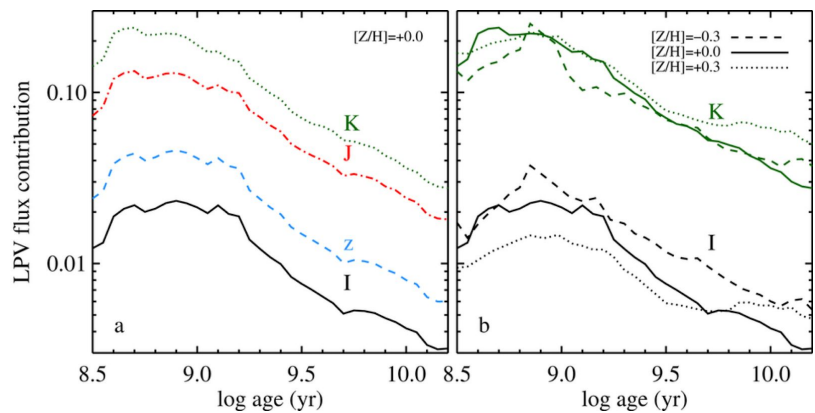

**Figure 1.** Fractional contribution of LPVs (TP-AGB stars) to total galaxy luminosity in various bands versus age. *Left:* solar metallicity models in *I*, *z*, *J*, and *K*. *Right:* comparison of half-solar, solar, and twice-solar ([Z/H] = −0.3, 0, +0.3) models in *I*, *K*. Both panels are reprinted from Conroy et al. (2015). Other models predict even higher LPV contributions at intermediate ages, by up to a factor of two.

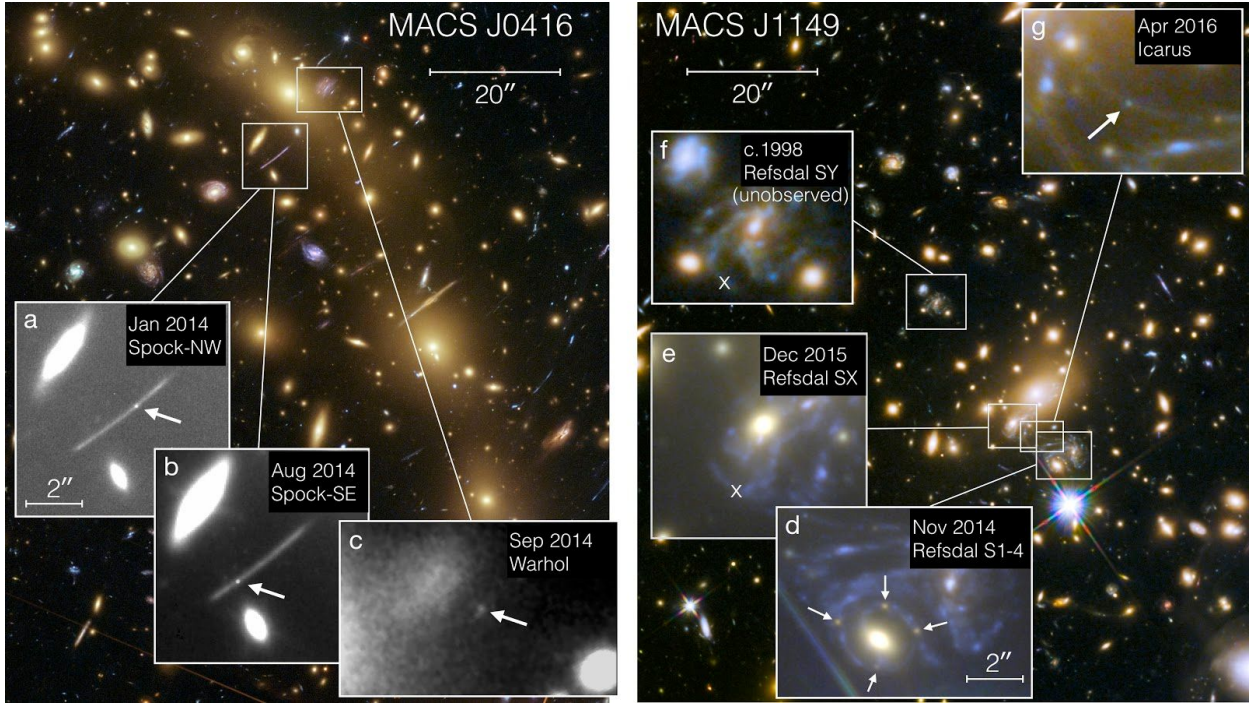

**Figure 2.** Strongly lensed transients discovered behind galaxy clusters MACS J0416 (left) and MACS J1149 (right). Wide-field high-resolution imaging of these clusters has been remarkably fruitful. In 2014, the MACSJ0416 cluster revealed a pair of peculiarly fast-evolving transients in a strongly-lensed galaxy (insets a and b). Nicknamed "Spock-NW" and "Spock-SE", these may have been strongly lensed outbursts of a massive star, or a pair of unrelated gravitational microlensing events (Rodney et al. 2018). In the follow-up imaging after the Spock-SE event, a new and unrelated transient appeared (inset c). Dubbed "Warhol," this event is interpreted as a single blue supergiant, magnified by a factor of 1000 or more (Chen et al. 2019). Not to be outdone, the MACS J1149 cluster produced the first strongly-lensed SN with resolved images (Kelly et al. 2015). Appearing first in an Einstein cross pattern (inset d), this "SN Refsdal" reappeared with a fifth observed image some 10" away (inset e). An earlier image, unobserved, would have been present across the cluster circa 1998 (inset f). Measurement of the SN Refsdal time delays (e.g., Rodney et al. 2016) could provide a constraint on H0 with ~7% precision (Grillo et al. 2018). Once again, SN Refsdal follow-up observations provided another discovery, with the appearance of the strongly-lensed star M1149-LS1 ("Icarus", inset g), the archetype for this new class of caustic-crossing transients due to the presence of a well-characterized persistent image (Kelly et al. 2018).

**Astrophysics and Cosmology of Gravitationally Lensed Transients**

The power of high angular resolution imaging for transient science has been demonstrated most dramatically in recent years by the Hubble Frontier Fields program (Lotz et al. 2017). This multi-year project invested 840 orbits with HST for multi-band NIR and optical imaging of 6 massive galaxy clusters, selected to be among the most efficient lenses on the sky. With supplementary observations from several related HST programs, the Frontier Fields clusters effectively became a showcase for the vast time-domain discovery space that is opened up by combining deep high-resolution imaging with strong gravitational lensing.

Figure 2 shows several of the most spectacular transient discoveries enabled by the Frontier Fields observations. All of these transients at $z \gtrsim 1$ are among the first of their class, and enabled studies touching on the astrophysics of massive stars (Kelly et al. 2016, Rodney et al. 2018), the distribution of cluster dark matter down to sub-halo scales (Kelly et al. 2018, Chen et al. 2019, Kaurov et al. 2019), and measurement of the cosmic expansion rate (Grillo et al. 2018, Vega-Ferrero et al. 2018). In most cases *these transients could not have been discovered without high angular resolution wide-field imaging, executed with a rapid cadence over a span of several months.* This combination has so far only ever been possible in the HST Frontier Fields. Importantly, the interpretation of these events requires precise lens models that span the entire cluster and incorporate intervening mass along the line of sight (Meneghetti et al. 2017, McCully et al. 2017). To build such models with accuracy and precision, it is critical to get high-resolution imaging over the entire cluster field (tens of arcseconds), and collect spectroscopic redshifts for dozens of galaxies within that field of view (Johnson & Sharon 2016).

The Frontier Fields time-domain effort can serve as a template for how to grow this nascent science field in the next decade. LSST imaging of strong-lensing clusters could provide early transient detections, and JWST imaging and spectroscopy could eventually deliver exquisite constraints for lens models. *The addition of a ground-based MCAO system with imaging and IFS capabilities would be a powerful bridge between the two, enabling rapid follow-up that JWST cannot do, and deep, high-resolution IR imaging and spectroscopy that LSST cannot provide.* If such a system had been available during the Frontier Fields program, it would have been the first choice for follow-up observations of all of the transients shown here.

**Further Opportunities**

A dynamically scheduled MCAO facility with a 2′ field, feeding an imager, MOS, and IFUs, would open many other opportunities for innovative "diffraction-limited time-domain" work. Examples span the range from time-series monitoring of weather and impacts on Jupiter and Saturn, to reverberation mapping of large samples of AGN to determine broad emission-line time delays to constrain dark energy (AO-assisted IFS allows probing lower luminosity AGN, which have shorter, more easily measured delays). For multiply-lensed QSOs, spectrophotometric monitoring can determine both the reverberation time delay to get the intrinsic luminosity (and thus magnification) and the lensing time delays, which together put tight constraints on $H_0$. Microlensing studies of the same systems also probe dark matter substructures within the lensing galaxies. In our own Galaxy, spectrophotometric monitoring with multiple AO-assisted IFUs (Sivanandam et al. 2018) covering a large patrol area to ensure precise relative calibration can be used to study clouds and weather on brown dwarfs. Other multi-epoch applications include precision astrometric studies of sources too faint or too crowded for GAIA, such as measuring the internal motions of stars in globular clusters to sufficient precision (~ 0.2 mas, with a multi-year baseline) to detect or rule out the presence of intermediate mass black holes.

Of course, such a system is also well-suited for many other high-resolution studies of distant and nearby galaxies, including much of the same science planned for JWST (see Kalirai 2018). Although no ground-based facility will compete with JWST in terms of background, a flexible system with rapid target acquisition could be very efficient for surveying multiple fields, especially in the case of resolved stellar populations. Eventually, TMT will provide 4x higher resolution with its NFIRAOS system, but over a 30″ field, or ¹⁄₁₆ the field size envisioned here. Further, an MCAO system combined with an adaptive secondary mirror (**ASM**) could be adapted for a much wider ~ 10′ field ground-layer AO (GLAO) mode (Hart et al. 2010) that would give a factor of 2-3 improvement in the PSF at all wavelengths with respect to the natural seeing. This would dramatically improve the efficiency for all scientific investigations focusing on unresolved sources, ranging from the solar system to cosmic dawn (e.g., ULTIMATE-Subaru Working Group 2016; Lu et al. 2018). Finally, we note that to achieve the full scientific benefits, these capabilities should be open to competitive access to the full astronomical community.

**Recommendation**

As discussed in the preceding sections, compelling science would be enabled by a facility that can deliver resolution and field-of-view comparable to JWST and that could also provide rapid time-domain follow-up and monitoring. This capability can be achieved with an advanced AO system operating on an 8m-class telescope, widely available to the US community. To maximize the science return, the characteristics of such a system should include:

- **Diffraction limited in *K* band**, giving FWHM ≈ 0.067″, with an imaging pixel scale matched to provide Nyquist sampling or better at this wavelength;
- **Corrected field-of-view at least 2′ in diameter**, to encompass the strong lensing regions of intermediate-redshift clusters and match approximately the size of JWST/NIRCam;
- **Nightly availability in a dynamically scheduled queue,** to enable time-domain studies ranging from rapid follow-up to long-term monitoring campaigns;
- **Sky coverage > 50% at the Galactic poles**, and virtually complete at low Galactic latitude, to ensure that most extragalactic fields of interest are observable;
- **Located at a premier astronomical site,** and on a telescope that **minimizes thermal background,** to maximize depth of observations;
- **Relative astrometric precision < 0.2 mas** across the 2′ field for studying the internal kinematics of dense, resolved stellar systems;
- **Operable at red optical wavelengths**, including *r, i, z* imaging and the ability to feed a multi-object spectrograph to optimize stellar population parameter diagnostics;
- **Wavelength coverage extending to 5 um** to maximize synergy with JWST, probe broad QSO emission lines such as Mg II out to z ~ 10, and take advantage of site conditions;
- **Multiplexed IFU capability,** patrolling the full corrected field, for detailed 2-D velocity maps of protostellar outflows, dense star clusters, and distant galaxies (requiring R~6000); also useful for isolating spectra associated with transients embedded in extended sources;
- **A wider-field, lower resolution mode,** provided by ground-layer AO correction, would facilitate a wide range of other science over a wider area; the ASM required for this mode would also reduce the complexity of the narrower-field AO system, thereby reducing the thermal background to maximize sensitivity.

All of these requirements are achievable with a cutting-edge MCAO facility utilizing sensitive wavefront sensors and combined with an adaptive secondary mirror on a modern, IR-optimized, queue-scheduled 8m-class telescope. Such a facility would provide superb synergy with JWST during extended periods when targets of interest are unobservable by JWST itself and in an era when Hubble may no longer be operational. It would also synergize with LSST, LIGO-Virgo, WFIRST, and other time-domain and multi-messenger surveys by enabling detailed photometric and spectroscopic monitoring of variable sources requiring high spatial resolution.